\documentclass[12pt]{article}

\usepackage{amsmath,amssymb}
\usepackage{graphicx}
\usepackage{caption}
\usepackage{hyperref} 
\usepackage{authblk}  
\usepackage[superscript]{cite}    
\usepackage{setspace} 
\usepackage{graphicx} 
\graphicspath{Images/}
\usepackage{booktabs,multirow} 
\usepackage{xr}
\usepackage{arydshln}
\usepackage{soul}
\usepackage{comment}
\usepackage{subcaption}
\usepackage{import}
\usepackage{float} 

\hypersetup{
   colorlinks=true,      
   linkcolor=blue,       
   citecolor=blue,       
   urlcolor=blue,        
   pdfpagemode=UseNone,  
}

\title{\textbf{{\fontfamily{ppl}\selectfont Accelerated Design of Microring Lasers with Multi-Objective Bayesian Optimization}\vspace{0.5em}}}

\author[*,1,4]{\fontfamily{cmss}\selectfont Mihir R. Athavale}
\author[1,2]{\fontfamily{cmss}\selectfont Ruqaiya Al-Abri}
\author[1]{\fontfamily{cmss}\selectfont Stephen Church}
\author[3]{\fontfamily{cmss}\selectfont Wei Wen Wong}
\author[4,5]{\fontfamily{cmss}\selectfont Andre KY Low}
\author[3]{\fontfamily{cmss}\selectfont Hark Hoe Tan}
\author[4,5]{\fontfamily{cmss}\selectfont Kedar Hippalgaonkar}
\author[*,1]{\fontfamily{cmss}\selectfont Patrick Parkinson}

\affil[1]{\normalsize \textit{Department of Physics and Astronomy and The Photon Science Institute, University of Manchester, Oxford Road, Manchester, M13 9PL, United Kingdom}}
\affil[2]{\normalsize \textit{College of Applied Sciences, University of Technology and Applied Sciences, Muscat 133, Sultanate of Oman}}
\affil[3]{\normalsize \textit{ARC Centre of Excellence for Transformative Meta-Optical Systems, Department of Electronic Materials Engineering, Research School of Physics, The Australian National University, Canberra ACT 2600, Australia}}
\affil[4]{\normalsize \textit{Institute of Materials Research and Engineering (IMRE), Agency for Science, Technology and Research (A$^{*}$STAR), 2 Fusionopolis Way, Innovis \#08-03, Singapore, 138634, Republic of Singapore}}
\affil[5]{\normalsize \textit{School of Materials Science and Engineering, Nanyang Technological University, Singapore, 639798, Singapore}\vspace{1em}}

\affil[ ]{ 
\small Email: \texttt{mihirrajendra.athavale@postgrad.manchester.ac.uk}; \,\ \texttt{ruqaiya.k.alabri@utas.edu.om}; \,\  \texttt{stephen.church@manchester.ac.uk}; \,\  \texttt{weiwen.wong@anu.edu.au}; \,\  \texttt{kaiyuana001@ntu.edu.sg}; \,\   \texttt{hoe.tan@anu.edu.au}; \,\   \texttt{kedar@ntu.edu.sg}; \,\   
\texttt{patrick.parkinson@manchester.ac.uk}}

\date{}

\newcommand{\thrE}{$189_{138}^{270}$\,$\mu$J cm$^{–2}$ pulse$^{–1}$}
\newcommand{\wvlE}{$1106_{1055}^{1146}$\,nm}
\newcommand{\thrF}{$270_{198}^{374}$\,$\mu$J cm$^{–2}$ pulse$^{–1}$}
\newcommand{\wvlF}{$957_{953}^{970}$\,nm}
\newcommand{\wvlO}{$1268_{1241}^{1286}$\,nm}

\begin{document}

\maketitle

\doublespacing

\begin{abstract}
On-chip coherent laser sources are crucial for the future of photonic integrated circuits, yet progress has been hindered by the complex interplay between material quality, device geometry, and performance metrics. We combine high-throughput characterization, statistical analysis, experimental design, and multi-objective Bayesian optimization to accelerate the design process for low-threshold, high-yield III-V microring lasers with room-temperature operation at communication wavelengths. We demonstrate a 1.6$\times$ reduction in threshold over expert-designed configurations, achieving a 100\,\% lasing yield that emits within the O-band with a median threshold as low as 33\,$\mu$J cm$^{-2}$ pulse$^{-1}$. 
\end{abstract}

\section*{Introduction}
Coherent light sources that can be directly integrated into photonic circuits, also known as on-chip lasers, represent one of the most challenging components to develop for future photonic integrated circuits (PICs) \cite{Past_to_future, Prosp_apps_lasers, Sun2015, Shen2017}. A variety of approaches have been proposed, including compound semiconductor PICs or heterogeneous integration via flip-chip (hybrid) techniques or monolithic heteroepitaxy \cite{Prosp_apps_lasers}. Explored gain architectures include semiconductor nanowires\cite{Bissinger2019,Schuster2017,Stettner2017, Kanungo2013, Parkinson_phys_apps_NWs}, quantum dots\cite{Shang2022ElectricallyWafers, Zhu2019ElectricallySi, Norman2018Perspective:Circuits} and, recently nanoridge diodes\cite{DeKoninck2023GaAsLine}.  

Bottom-up grown III-V multi-quantum well (MQW) microring (MR) lasers present an intriguing structure due to their wavelength tunability through quantum confinement, and potential for efficient coupling to in-plane waveguides, making them promising candidates in next-generation PICs \cite{Wong2021EpitaxiallyLasers,Wong_bottom_up_MQW_MRs}. In common with many bottom-up structures, their performance is significantly influenced by fabrication challenges, such as complex electron beam lithography (EBL) patterning, the need for precise control over growth rates across multiple facets, as well as a range of other optical, geometrical, and material parameters \cite{Wong_bottom_up_MQW_MRs}. Managing these factors which are sensitive to variations far below the wavelength scale poses significant reproducibility challenges \cite{Al-Abri2021_ACS}.

To address these complexities, there is an increasing focus on applying rigorous statistical methods such as experimental data driven discovery to optimize such novel devices \cite{Church2022HolisticNanowires, Church2023HolisticDesign, Church_data_driven_discovery_2024}. Yet, simultaneously achieving high yield, low lasing thresholds, and precisely tuned lasing wavelengths remains a complex multi-objective challenge, especially when quantum-confined heterostructures are involved \cite{Skalsky2020_ACS}.
In this study, we present a high-throughput methodology for developing efficient (low threshold), reliable (high yield), and targeted (O-band communication wavelength --- between 1260 and 1360 nm) room-temperature, multi-mode lasing MQW MR lasers. Our approach uses Multi-Objective Bayesian Optimization (MOBO) which leverages surrogate models such as Gaussian process regression to build a probabilistic representations of the objectives. We use this function to create optimized designs, and augment these with a Design of Experiments (DoE) approach to efficiency explore the parameter space.

Our results demonstrate that all groups (corresponding to separate \textit{fields}) of nominally-identical optimized MQW MR lasers achieved a median lasing threshold of less than 95\,$\mu$J cm$^{-2}$ per pulse. The process achieved a 100\,\% yield in 25 out of 34 fields grown with optimized parameters, with every field showing a yield exceeding 96\,\%. This successful result was achieved in a single growth cycle, underscoring the efficiency of our approach. We observed a $\sim65$\,\% reduction in lasing thresholds for the best-performing field compared to the best previously reported values for InP/InAsP MQW MRs fabricated using conventional (expert-designed) methods \cite{Wong_bottom_up_MQW_MRs}. Lasing was observed at communication wavelengths, with the best-performing field exhibiting a median lasing wavelength in the O-band (\wvlO - interquartile range denoting the upper and lower limits). A total of 251 MQW MRs out of around 2000 exhibited lasing in the O-band, each with a threshold of less than 50\,$\mu$J cm$^{-2}$ pulse$^{-1}$. These results highlight the potential of our approach which can serve as an additional tool to help epitaxial growers to navigate a multi-dimensional optimization space.

\section*{Results}
\subsection*{Training Dataset : Before Optimization}

\begin{figure}[ht]
\centering
\includegraphics[width=1\textwidth]
{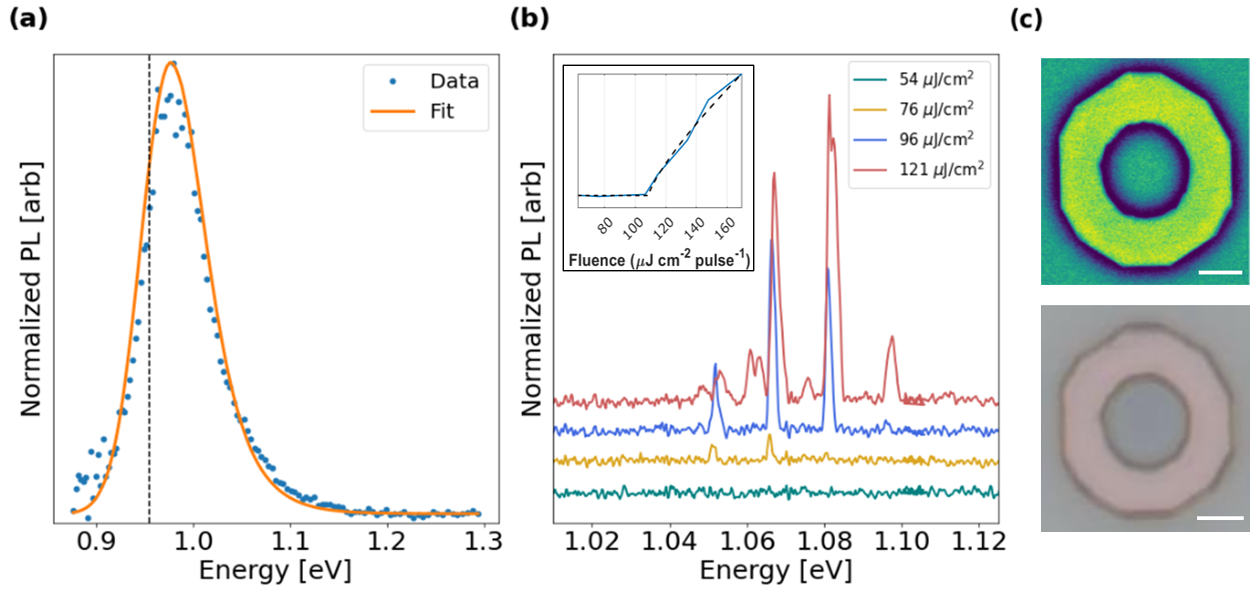}
\caption{Spectroscopy and imaging for a single selected MQW MR (sample F, field 9, ring number 647). (a) Room-temperature photoluminescence (PL) spectrum acquired at low excitation power (below the lasing threshold), with the convoluted Boltzmann-density of states fit shown as a solid line. The vertical line indicates the MQW band-edge obtained from the fit. (b) Evolution of emission spectra with excitation fluence varying from 54\,$\mu$J cm$^{-2}$ pulse$^{-1}$ to 121\,$\mu$J cm$^{-2}$ pulse$^{-1}$ (offset for clarity). The inset shows an integrated light-in light-out (LILO) curve representing intensity versus fluence (in $\mu$J cm$^{-2}$ pulse$^{-1}$), with a fit (dashed line) used to determine the lasing threshold. (c) (Top) False-colored scanning electron microscopy (SEM) image and (bottom) optical image of the MR, with a 2\,$\mu$m scale bar.
}
\label{fig:MR_PL_images}
\end{figure}

To generate an initial training set for optimizing the growth parameters and geometries of MQW MRs, six samples were fabricated using selective area epitaxy (SAE) via metal-organic chemical vapour deposition (MOCVD), following an established approach \cite{Yuan2021_Selective}. Each MQW MR comprises an InP core, alternating layers of InAsP/InP quantum wells (QWs) and barriers, as well as an InP capping layer, all grown on an InP substrate. Comprehensive growth details are available in a recent study \cite{Wong_bottom_up_MQW_MRs}. Each sample was produced using distinct growth recipes, with nine fields of MQW MRs per sample. These fields contained between 48 and 120 nominally identical MQW MRs, with different fields having variations in geometrical parameters, specifically diameter ($D$) and pitch ($P$)—the distance between the centers of neighboring MQW MRs. The selection of growth recipes and geometries for the initial set was guided by domain experts, and the growth and geometry parameters used for subsequent analysis are summarized in Table \ref{tab:training}.

The MQW MRs were identified and analyzed using automated micro-photoluminescence ($\mu$-PL) spectroscopy, employing a machine vision approach \cite{Church2022HolisticNanowires, Parkinson2020}. Initially, low-power PL measurements were performed at room temperature. Figure \ref{fig:MR_PL_images} (a) shows an example of the PL spectrum for a single MR, with emission attributed to transitions in the MQW. To determine the lasing threshold and wavelength of each MR laser, power-dependent PL measurements were conducted at room temperature, following the previously introduced approach \cite{Alanis2017,Church2023HolisticDesign, Wong_bottom_up_MQW_MRs}. An example of power-dependent measurements for a single MQW MR is shown in Figure \ref{fig:MR_PL_images} (b), illustrating the evolution of PL spectra with increasing fluence from 54 to 121\,$\mu$J cm$^{-2}$ pulse$^{-1}$. The spectra show multi-mode lasing, with modes spanning 1.04 to 1.11\,eV. The lasing thresholds were calculated from the intensity of the dominant lasing emission peak.

For each field, the median values of the lasing threshold and wavelength were calculated. The yield was defined as the ratio of MRs that exhibited lasing at or below the maximum fluence used (850\,$\mu$J cm$^{-2}$ pulse$^{-1}$) to the total number of MRs in that field. In total, 4,128 MQW MRs were analyzed across 53 combinations of six growth recipes (samples) and nine geometrical configurations. These were mapped to three performance metrics: lasing threshold, lasing wavelength and yield. This data was used to form a training dataset. 

For optimization, we considered only five relevant growth parameters (see SI Table S1). This resulted in a total of seven distinct features: five growth parameters ($nQWs$ - number of quantum wells , $T_g$ - growth temperature, $As/P$ - ratio of As to P ratio calculated using the flow rates of the precursors during the QW growth (in vapour phase), $P/In$ - V/III ratio during InP barrier growth , and $t_g$ - growth time of the capping layer) and two geometry parameters ($D$ and $P$). Table 1 in the Methods section presents the resultant tables of distinct, variable, and optimizable features.

\begin{figure}
\centering
\includegraphics[width=\textwidth, height=6.7cm]{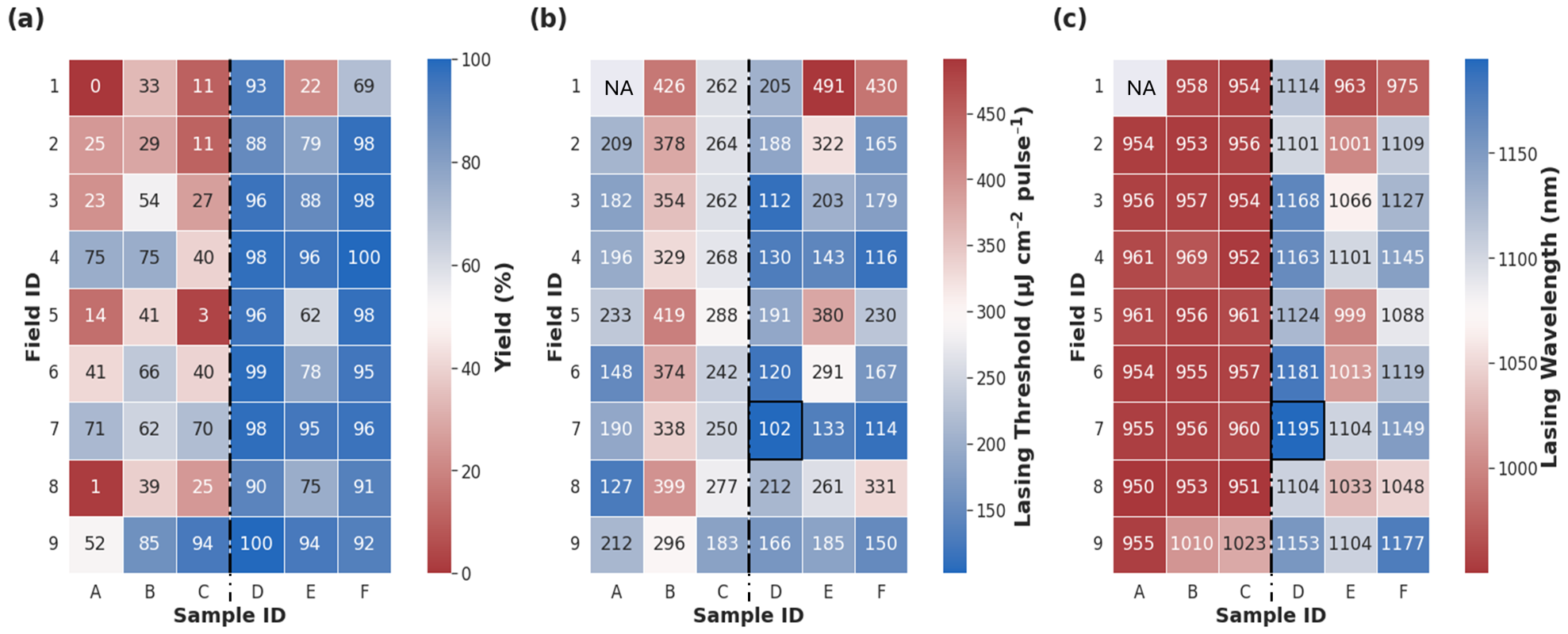}
\caption{Heat maps illustrating the variations in (a) yield, (b) field-median lasing thresholds, and (c) field-median lasing wavelengths from the training dataset. The top-performing fields for lasing threshold and lasing wavelength are outlined with squares. A vertical line separates samples with 5 (A,B,C) and 8 (D,E,F) QWs.}
\label{fig:heatmap_MR_lasing_initial}
\end{figure}

In Figure \ref{fig:heatmap_MR_lasing_initial} (a), the yield percentages for all samples across different fields are presented. The highest yield of 100\,\% is observed in samples with 8 QW in the fields with the largest pitch and diameter(\textbf{D\#9}, \textbf{F\#4} where  \textbf{D, F} are the sample identifiers, and  \textbf{\#9, \#4} are the field numbers). Figures \ref{fig:heatmap_MR_lasing_initial} (b) and (c) provide maps of the lasing threshold and lasing wavelength, respectively. The minimum median lasing threshold, 102\,$\mu$J cm$^{-2}$ pulse$^{-1}$, is achieved in sample D\#7, which also exhibits the longest lasing wavelength of 1195\,nm. These results highlight a median lasing threshold of \thrE~ for the 8\,QW samples, significantly lower than the \thrF~ observed for the 5\,QW samples, with the interquartile range denoting the upper and lower limits. Furthermore, the median lasing wavelength for the 8\,QW samples is \wvlE~, compared to \wvlF~ for the 5\,QW samples, suggesting superior lasing performance in the 8\,QW structures.

The enhanced performance with an increased \textit{nQWs} can be attributed to improved overlap between the spatial overlap between the lasing modes and the active region \cite{Stettner2016}. However, it is important to consider the influence of other growth factors on this improvement. One approach to identifying the most critical parameters is Univariate Feature Selection (UFS), which selects key features through univariate statistical tests including chi-squared, Analysis of Variance (ANOVA), and Mutual Information \cite{UFS_Abellana_2023}. A UFS analysis reveals that $nQWs$, $As/P$, and $T_g$ are the three most important parameters across all targets (more details on UFS are provided in the SI). While UFS can give information about the dependence of targets on features, it is not well suited to predicting optimal features where there is a non-linear relationship between feature and target. 

We use two design approaches to develop optimal growth recipes -- MOBO and DoE. These were selected due to the constraints of our problem. Firstly, our three chosen objectives are a mix of minimization and maximization targets: a low lasing threshold (minimize), a finely tuned lasing wavelength for telecommunication (maximize), and high fabrication yield (maximize). Secondly, growth parameters in our training set show limited variation; for example only two distinct $nQWs$, 5 and 8, leading to significant gaps (Table \ref{tab:training}). Finally, the high costs and lengthy fabrication times involved in the manufacturing process limit the number of experimental runs available for optimization. MOBO was used to find optimal features that provide the best trade-offs between the desired objectives, known as \textit{non-dominated} or \textit{Pareto-optimal} solutions \cite{Low_mapping_pareto} while Principal Component Analysis with Design of Experiments (PCA-DoE) is used to fill gaps in the dataset and facilitate further exploration. In this approach, we used  PCA to reduce the dimensionality of the growth parameters into one axis, and reduce the geometry parameters into another. This approach reduces the design space to two dimensions, allowing a scatter plot to be used to effectively navigate the search space (see more in the SI). PCA was only used for the DoE approach; a detailed descriptions of both techniques can be found in the Methods section. 

\subsection*{Optimized \& DoE Dataset}

We used MOBO to generate four optimized growth parameters (\textbf{G, H, I, J}), and PCA-DoE to generate one additional exploratory sample (\textbf{DoE}). A summary of the new geometries and growth conditions is provided in Table 2 in the Methods section. Field 1 in sample \textbf{G} was excluded from measurements due to failed ring growth caused by faulty mask openings. Additionally, only seven fields with varying geometries were fabricated in the DoE sample, resulting in 42 new combinations.

\begin{figure}[ht]
\centering
\includegraphics[width=\textwidth, height=6.7cm]{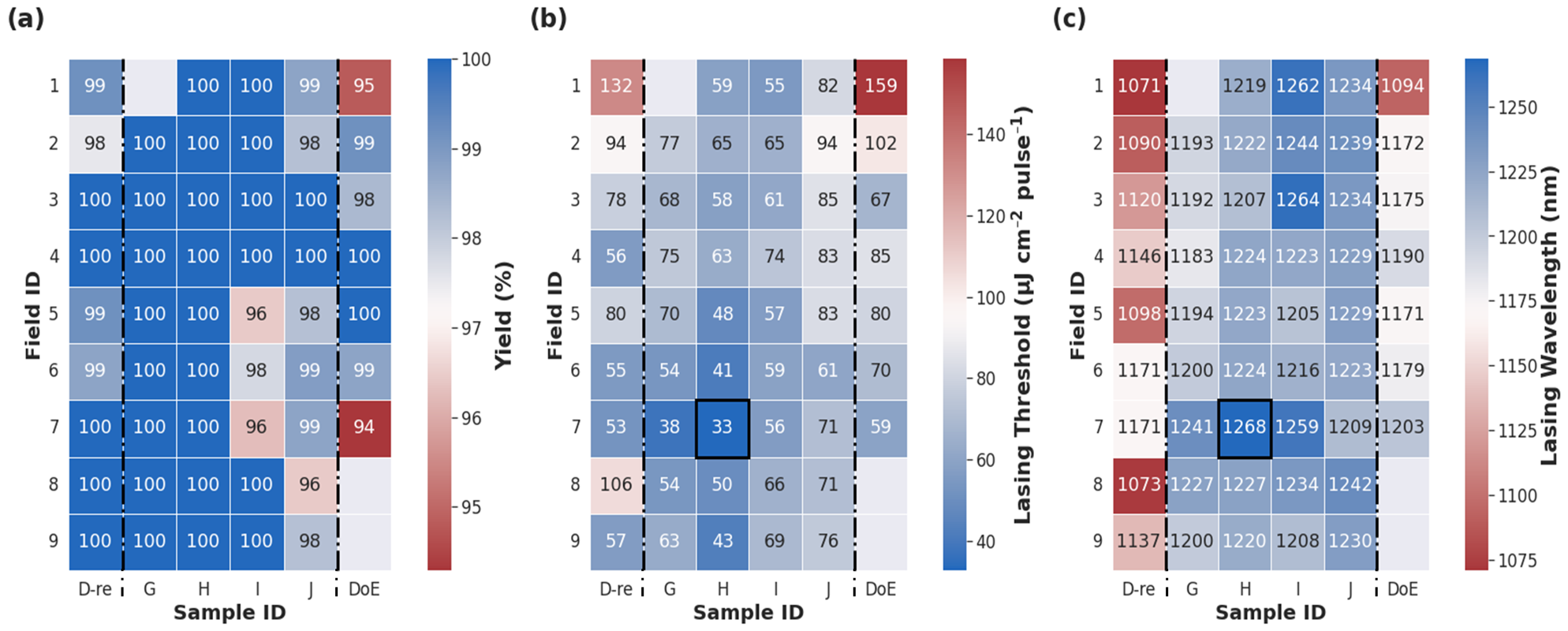}
\caption{Heat maps illustrating variations in (a) yield, (b) field-median lasing thresholds, and (c) field-median lasing wavelengths between a remeasured previous best sample (\textbf{D})
and optimized samples including DoE. A black square is used to outline the best performing field (\textbf{H\#7}).}
\label{fig:Heatmap_MR_lasing_optimized}
\end{figure}

Measurements of the optimized and DoE MQW MR samples were carried out under the same calibration settings and nominal experimental conditions as those used for the training dataset. Additionally, the best performing sample was remeasured as an additional check for comparability (shown as \textbf{D-re}). We note that remeasurement provided a change in sample \textbf{D} performance -- a 3\% blue-shift in wavelength and a 50\% decrease in threshold (see more detail in the SI). The former is likely insignificant given the 20\% standard deviation across the sample. The latter is significant, but attributable to improvements in excitation beam profile and alignment between measurement runs. We therefore reference all improvements to the remeasured sample for consistency and robustness.

\begin{figure}[H]
\centering
\includegraphics[width=0.9\textwidth, height=11.8cm]{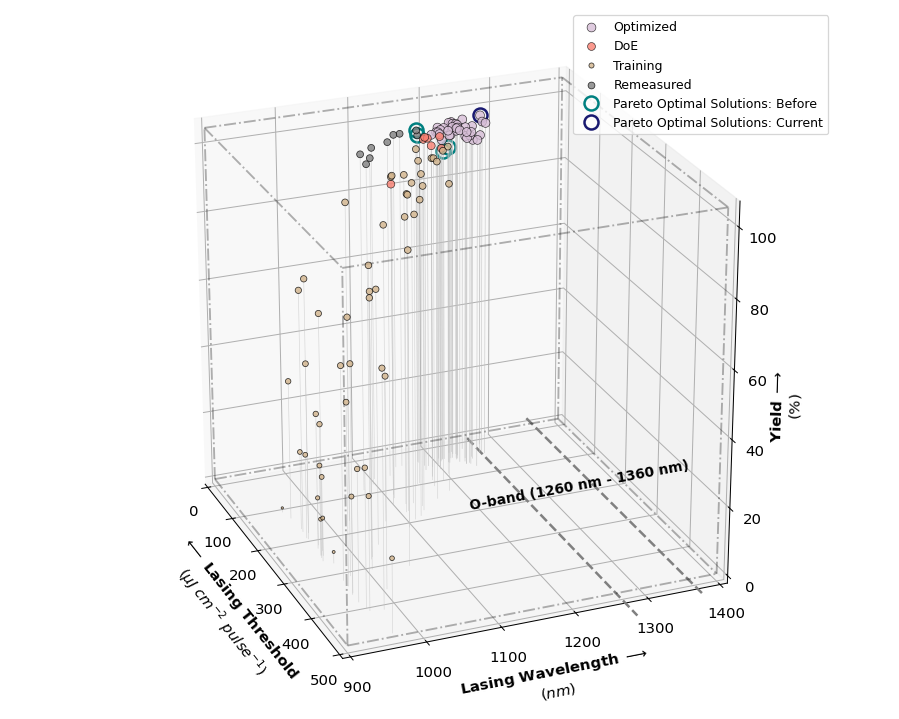}
\caption{
Scatter plot illustrating three measured performance metrics -- median lasing threshold ($\mu$J cm$^{-2}$ pulse$^{-1}$), median lasing wavelength (nm), and yield (\%) -- for every measured field. The color of the points indicates before/after optimization (yellow and purple, respectively), with the remeasured best-performing sample (in grey), as well as those from the DoE growth (in red).
The Pareto-optimal solutions are denoted with a thick outer circle. The optimized features demonstrate improvement over previous Pareto-optimal solutions.}
\label{fig:Obj_scatter_plot_combined}
\end{figure}

Figure \ref{fig:Heatmap_MR_lasing_optimized} (a) reveals that samples \textbf{G} and \textbf{H} have a 100 \,\% yield across all fields. In the optimized dataset, no field exhibited a yield lower than 96\,\%. Improvements were observed in other performance metrics: the median lasing threshold remained consistently low, below 100 $\mu$J cm$^{-2}$ pulse$^{-1}$ across all optimized fields, and the median lasing wavelengths were consistently at or above the previous best value of 1195 nm, with three fields having median lasing wavelengths in the telecommunications O-band.  This achievement marks an improvement over previously reported best values ($\sim$90\,\% yield, thresholds of $\sim$100 $\mu$J cm$^{-2}$ pulse$^{-1}$ \cite{Wong_bottom_up_MQW_MRs}). The optimized samples also outperformed those from both the exploratory \textbf{DoE} dataset and the training dataset. Among the optimized samples, \textbf{H} emerged as the top performer, with Field 7 within this sample demonstrating exceptional results: a 100\,\% yield, a median threshold of 33 $\mu$J cm$^{-2}$ pulse$^{-1}$, and a median lasing wavelength of 1268 nm, which falls within the communication O-band. This represents a simultaneous $1.6\times$ improvement in threshold and a 90\,nm redshift in lasing over the previous champion field \textbf{D-re\#7} measured under the identical conditions, in a single optimization iteration. Figure \ref{fig:Obj_scatter_plot_combined} illustrates a comparative analysis of the measured objectives, highlighting changes before and after the optimization process. Improvements are achieved through optimization, with all output metrics outperforming or equalling the initial Pareto-optimal solutions. 

\section*{Discussion}
Continuous-wave, telecommunication-range, low-threshold lasers are essential for enhancing commercial on-chip light sources in PICs. Our approach marks a significant advancement toward this objective, providing an immediate $1.6\times$ reduction in threshold, achieving 100\,\% yield rates, and tuning wavelength performance in InP/InAsP MQW MR lasers in a single optimization cycle. We have demonstrated that a precisely controlled set of parameters can be used to develop a surrogate model with predictive capabilities that can outperform traditional expert methods. This achievement is particularly noteworthy given the challenges posed by the inherent variability in both measurements and fabrication due to fluctuating external environmental conditions.

This success highlights the potential of our methodology to leverage existing datasets for optimizing future growth processes, particularly in developing complex structures with non-linear performance metrics. This approach provides expert growers with an additional tool for navigating the high-dimensional search space to achieve optimal outcomes while also opening promising opportunities for workflow automation in such processes. Nevertheless, there are areas for further refinement in our methodology. Currently, our optimization focuses primarily on median thresholds and wavelengths, without addressing the inhomogeneity and ring-to-ring variance within fields. Future work should aim to minimize this variance, achieving a more uniform set of MQW MRs capable of lasing within the communication wavelength range while maintaining low thresholds.


\section*{Materials and Methods}
\label{section:Methods}
\subsection*{Experimental Methods}

\paragraph{Low-Power Photoluminescence (PL)}
 A continuous-wave 532\,nm laser was focused on the MQW MR, with a power density at the sample of 170\,W cm$^{-2}$.  Each PL spectrum was fitted using a convoluted Boltzmann-Density of States model, described previously \cite{Al-Abri2023Sub-PicosecondFramework}, which provides the transition energy, intensity, disorder broadening, and effective emission temperature.

\paragraph{Lasing}
An excitation laser pulse, with a duration of approximately 200\,fs, wavelength of 633\,nm, and a repetition rate of 100\,kHz, was utilized. The light was directed onto the sample through a 20$\times$ magnification, 0.75 NA objective lens. A defocusing lens facilitated nearly uniform excitation across a spot size of 37\,$\mu$m in diameter. Power-dependent PL measurements were conducted at room temperature by adjusting the incident fluence up to 850\,$\mu$J cm$^{-2}$ pulse$^{-1}$ using a neutral-density filter wheel. The collection of power-dependent spectra was halted when a decrease in emission was observed or until the incident fluence limit (850 \,$\mu$J cm$^{-2}$ pulse$^{-1}$), and measurements were not extended to the damage threshold, allowing for remeasurement. The emission was captured with the same objective lens, spectrally filtered to remove the excitation light, and focused onto a 100\,$\mu$m diameter optical fiber with a numerical aperture (NA) of 0.22. This fiber was then connected to a Horiba iHR550 spectrometer, with a slit width of 0.1\,mm. The spectrum was recorded using a 150\,lines/mm grating and an Andor iDus InGaAs detector array.

\begin{table}[ht]
\centering
\caption{Summary of the training set of features consisting of growth parameters and MQW MR geometries used for optimization. (a) Growth parameters include the number of quantum wells (nQWs), growth temperature (T$_g$), the ratio of As to P in the quantum well (As/P), the V/III ratio during InP barrier growth (P/In), and the growth time of the capping layer (t$_g$). (b) MQW MR geometries include the diameter (D) and pitch (P). The column indicating the number of MRs is provided for reference; however, this is not a feature used for optimization.\label{tab:training}}
\begin{tabular}{ccccccc}
\hline
\multicolumn{6}{c}{\textbf{(a) Growth Parameters}} \\
\hline
Sample & nQWs & T$_g$ [$^0$C] & As/P [$\%$] & P/In  & t$_g$ [s] \\
\hline
A & 5  & 600 & 0.66 & 2000 & 120 \\
B & 5  & 600 & 1.04 & 2000 & 180 \\
C & 5  & 590 & 0.79 & 2000 & 180 \\
D & 8  & 590 & 0.92 & 2000 & 180 \\
E & 8  & 600 & 0.92 & 2000 & 180 \\
F & 8  & 600 & 0.92 & 1500 & 240 \\
\hline
\end{tabular}

\vspace{0.5cm} 

\begin{tabular}{cccc}
\hline
\multicolumn{4}{c}{\textbf{(b) Microring Geometries}} \\
\hline
Field & D [$\mu$m] & P [$\mu$m] & \textit{Number of MRs} \\
\hline
1 & 4 & 7.5  &  \textit{120}  \\
2 & 6  & 10  &  \textit{80}  \\
3 & 8  & 12  &  \textit{56}  \\
4 & 10  & 13 &  \textit{48}  \\
5 & 5 & 7.5  &  \textit{120} \\
6 &  7 & 10  &  \textit{80}  \\
7 &  9 & 12  &  \textit{56}  \\
8 &  5 & 10  &  \textit{80}  \\
9 &  7 & 14  &  \textit{48}  \\
\hline
\end{tabular}
\label{tab:features}
\end{table}

\paragraph{Geometry - Imaging}
Selective area growth generally provides better control of geometric uniformity than vapor-liquid-solid methods \cite{Zhang2021AApplications,Rudolph2014,Noborisaka2005}. However, heterostructural growth tends to increase inhomogeneity \cite{Wong_bottom_up_MQW_MRs}. Given the complex, close relationship between geometry and performance, this is expected to result in inhomogeneous thresholds and potentially reduce yield. We used two imaging methods -- Scanning Electron Microscopy (SEM) and optical imaging -- for all MQW MRs to study the geometry. More details on imaging are given in SI.

\subsection*{Multi-objective Bayesian Optimization}

Multi-objective Bayesian Optimization (MOBO) is a global optimization technique designed to solve problems involving multiple, often conflicting objectives, making it particularly useful for optimizing expensive black-box functions where the underlying function is unknown and costly to evaluate \cite{qNEHVI_qNParEGO, Shahriari2016, Low_mapping_pareto}. It leverages surrogate models, such as Gaussian process regressors \cite{GP_Ramussen}, to create probabilistic representations of the objective functions and employs acquisition functions to balance exploration (searching unexplored areas) and exploitation (focusing on known promising regions) \cite{Shahriari2016}. This approach enables efficient navigation of the search space, guiding the optimization process toward optimal trade-offs across multiple objectives known as Pareto front \cite{Low_mapping_pareto}.

\begin{figure}[htbp]
\centering
\includegraphics[width=\textwidth, height=5.7 cm]{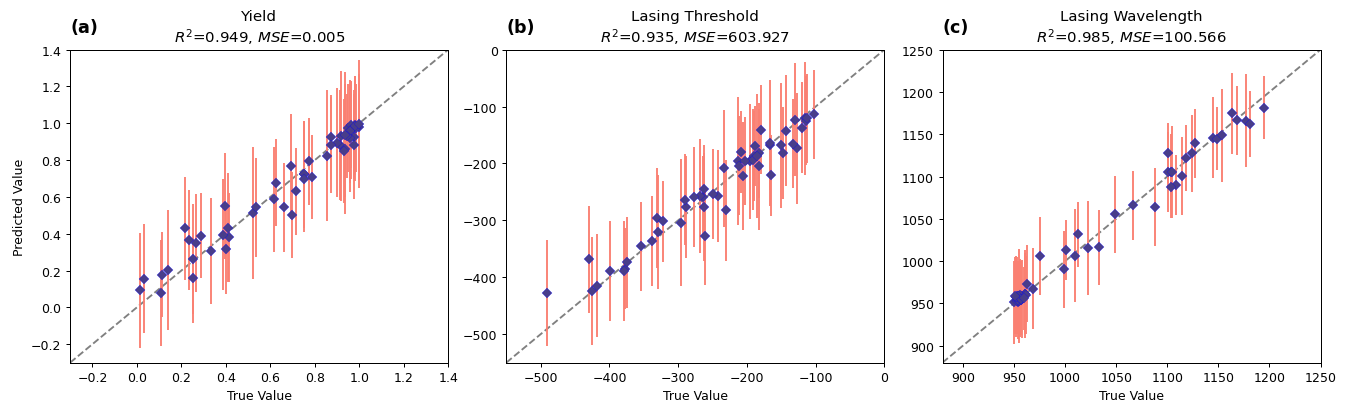}
\caption{Model accuracy of the Gaussian process regressor for (a) yield, (b) field-median lasing thresholds (note, values are inverted to change a minimization problem into a maximization objective), and (c) field-median lasing wavelengths, based on 53 training data points. Mean Squared Error (MSE) and R$^2$ scores are reported for each objective.}
\label{fig:GP_regression}
\end{figure}

The SingleTaskGP model in BoTorch \cite{BoTorch} was used to model the objectives. Figure \ref{fig:GP_regression} illustrates the model accuracy for all three objectives. Since the goal was to minimize the threshold, and BoTorch assumes maximization for all objectives, the lasing threshold values were negated before modeling. For the acquisition process, several commonly used algorithms were considered, including:

\begin{enumerate}
    \item \textbf{q-Noisy Expected Hypervolume Improvement (qNEHVI)}, Hypervolume is a measure of the volume in the objective space that is dominated by Pareto-optimal solutions \cite{Low_mapping_pareto}. In this approach, solutions are evaluated based on how they contribute to the ``hypervolume" covered by the Pareto front. This provides a way to estimate how a new candidate solution could contribute in expanding or improving the Pareto front in a noisy setting \cite{qNEHVI_qNParEGO}.
    \item \textbf{Evolution Guided Bayesian Optimization (EGBO)}, which applies a hybrid framework combining Evolutionary algorithms with qNEHVI to effectively guide the optimization process \cite{EGBO}.
    \item \textbf{q-Noisy Pareto Efficient Global Optimization (qNParEGO)} , which uses the q-Noisy Expected Improvement (qNEI) acquisition function. This function transforms multiple objectives into a single objective through augmented Chebyshev scalarization \cite{qNEHVI_qNParEGO, ParEGO}, enabling more efficient optimization in certain contexts such as for example, when optimizing more than five objectives, where hypervolume-based methods tend to struggle \cite{ Lopez_Evolutionary_IGD_EMOA}.
\end{enumerate}

The limited number of Pareto-optimal solutions in our training set has resulted in a sparse Pareto front. Furthermore, there is no substantial Pearson correlation (greater than 0.7) among the objectives, with the exception of the correlation observed between lasing wavelength and yield (SI Figure S2). Consequently, qNParEGO, implemented in BoTorch, was selected as the preferred method, despite its potentially slower convergence compared to alternatives like qNEHVI \cite{qNEHVI_qNParEGO, Daulton_qEHVI}. This approach, however, generates diverse random combinations of objectives as output candidates, facilitating the exploration of varied configurations within a single iteration.

\begin{table}[ht]
    \centering
    \caption{Lower and upper bounds for growth and geometry parameters representing experimental constraints on each feature.}
    \begin{tabular}{c c c}
        \hline
        \textbf{Features} & \textbf{Lower Bound} & \textbf{Upper Bound}  \\
        \hline
        nQWs            &  1    &  12    \\
        T${_g}$[$^0$C]  &  590  &  620     \\
        As/P $\%$]      &  0.6  &  1.1   \\
        P/In            &  1500 &  4000      \\
        t$_g$ [s]       &  60   & 240      \\
        \hdashline
        D [$\mu$m]      &  4    & 15       \\
        P [$\mu$m]      &  6    & 30       \\
        \hline
    \end{tabular}
    \label{table:bounds}
\end{table}

The study focuses on a seven-dimensional feature space, consisting of five growth parameters (samples) and two geometry parameters (fields). For each set of growth parameters, nine different sets of geometry parameters were available. Using MOBO, four optimal sets of these seven-dimensional features were identified within the specified problem bounds outlined in Table \ref{table:bounds}. These bounds represent the experimental constraints on each feature, which are either determined by experts or reflect instrumental limitations. To further explore the impact of geometry, Latin Hypercube Sampling (LHS) \cite{McKay_LHS} was applied to generate a range of geometry parameter combinations around the optimized values for each set of growth parameters. This approach allowed for the examination of how variations in geometry parameters influence the performance of the optimized growth parameter sets. A summary of the optimized growth and geometry parameters, along with a DoE dataset, is provided in Table \ref{tab:optimized_DoE}.

\begin{table}[ht]
\centering
\caption{Summary of growth parameters and MQW MR geometries for optimized and DoE Datasets. (a) Growth parameters follow the same conventions as described in Table 1(a). The sample labeled `DoE-8' refers to the DoE sample, while the `D-8' sample, shown in italics, represents the best-performing sample from the training dataset. This comparison highlights the variations between the top-performing growth parameters of the training dataset and those of the optimized and DoE datasets. (b1) and (b2) present MQW MR geometries (D and P) for the optimized and DoE datasets, respectively, including the number of MRs per field, consistent with the details in Table 1(b).\label{tab:optimized_DoE}}

\begin{tabular}{c}
\begin{tabular}{ccccccc}
\hline
\multicolumn{6}{c}{\textbf{(a) Growth Parameters}} \\
\hline
Sample & nQWs & T$_g$ [$^0$C] & As/P [$\%$] & P/In & t$_g$ [s] \\
\hline
\textit{D} & \textit{8}  & \textit{590} & \textit{0.92} & \textit{2000} & \textit{180} \\
G & 9  & 595 & 0.88 & 1612 & 212 \\
H & 8  & 596 & 0.91 & 1643 & 225 \\
I & 9  & 594 & 0.92 & 1669 & 221 \\
J & 9  & 590 & 0.92 & 1846 & 198 \\
DoE & 8  & 598 & 1.10 & 1715 & 243 \\
\hline
\end{tabular}
\end{tabular}

\vspace{0.5cm} 

\begin{tabular}{cc}
\resizebox{0.5\textwidth}{!}{
\begin{tabular}{cccc}
\hline
\multicolumn{4}{c}{\textbf{(b1) Microring Geometries (Optimized)}} \\
\hline
Field & D ($\mu$m) & P ($\mu$m) & \textit{Number of MRs} \\
\hline
1 & 9 &  11.70  & \textit{80} \\
2 & 9  & 14.50  & \textit{56}\\
3 & 9  & 17.10  & \textit{56}\\
4 & 9  & 14.90  &  \textit{48}\\
5 & 9  & 16.51  & \textit{56}\\
6 & 9  & 13.48  & \textit{90}\\
7 & 10 & 15.20  & \textit{80}\\
8 & 10 & 16.46  & \textit{56 }\\
9 & 10 & 13.03  & \textit{56}\\
\hline
\end{tabular}
}
&
\resizebox{0.46\textwidth}{!}{
\begin{tabular}{lccc}
\hline
\multicolumn{4}{c}{\textbf{(b2) Microring Geometries (DoE)}} \\
\hline
Field & D ($\mu$m) & P ($\mu$m) & \textit{Number of MRs} \\
\hline
1 &  5  & 9.2  & \textit{154} \\
2 &  6  & 10   & \textit{120}\\
3 &  7  & 10.8 & \textit{120 } \\
4 &  8  & 11.7 & \textit{90}\\
5 &  9  & 12.6 & \textit{80}\\
6 &  10 & 13.4 & \textit{80}\\
7 &  11 & 14.3 & \textit{70}\\
\hline
\end{tabular}
}
\end{tabular}

\end{table}

\subsection*{PCA based Design of Experiments}
Principal Component Analysis (PCA) is a widely used technique for dimensionality reduction \cite{PCA_dim_red}. Significant gaps existed in the training dataset, with large unexplored regions between the available constraints on growth parameters (problem bounds). Given the limited number of growth samples, PCA was employed to effectively select a set of growth as well as geometry parameters for exploration within the specified bounds.
The growth parameters from the training dataset were transformed into a single principal component, while the geometry parameters were also converted into one principal component. A scatter plot of these principal components indicated that yield increased along the growth parameter axis from right to left (see SI Figure S6). Consequently, a strategic point was selected near one of the extreme bounds on the growth parameter principal axis.
To identify corresponding geometry parameter sets for this chosen point, LHS was applied along the geometry parameter axis. Transforming these sampled points back into the original growth parameter space generated a DoE set, which filled in the existing gaps and enabled a more comprehensive exploration of parameter space, ultimately enhancing the model predictive capabilities for future optimizations.

\section*{Data Availability}
The data supporting the findings of this paper is accessible at Figshare DOI: \href{https://dx.doi.org/10.48420/27330528}{10.48420/27330528}

\section*{Code Availability}
The code used in this study is available on GitHub at: \href{https://github.com/OMS-lab/Microring_MOBO}{https://github.com/OMS-lab/Microring\_MOBO}

\section*{Acknowledgements}
Patrick Parkinson acknowledges funding under the UKRI Future Leaders Fellowship scheme [MR/T021519/1] and EPSRC grant [EP/V036343/1]. Ruqaiya Al-Abri acknowledges the studentship funding from the Ministry of Higher Eduction, Research, and Innovation, Oman. The Australian authors acknowledge the Australian Research Council for financial support and The Australian National Fabrication Facility, ACT Node for access to epitaxial growth and device fabrication facilities. 

\bibliographystyle{naturemag.bst}  
\bibliography{references.bib}  

\begin{thebibliography}{10}
\expandafter\ifx\csname url\endcsname\relax
  \def\url#1{\texttt{#1}}\fi
\expandafter\ifx\csname urlprefix\endcsname\relax\def\urlprefix{URL }\fi
\providecommand{\bibinfo}[2]{#2}
\providecommand{\eprint}[2][]{\url{#2}}

\bibitem{Past_to_future}
\bibinfo{author}{Yang, J.}, \bibinfo{author}{Tang, M.}, \bibinfo{author}{Chen,
  S.} \& \bibinfo{author}{Liu, H.}
\newblock \bibinfo{title}{{From past to future: on-chip laser sources for
  photonic integrated circuits}}.
\newblock \emph{\bibinfo{journal}{Light: Science {\&} Applications}}
  \textbf{\bibinfo{volume}{12}}, \bibinfo{pages}{1--3} (\bibinfo{year}{2023}).

\bibitem{Prosp_apps_lasers}
\bibinfo{author}{Zhou, Z.} \emph{et~al.}
\newblock \bibinfo{title}{{Prospects and applications of on-chip lasers}}.
\newblock \emph{\bibinfo{journal}{eLight}} \textbf{\bibinfo{volume}{3}},
  \bibinfo{pages}{1--25} (\bibinfo{year}{2023}).

\bibitem{Sun2015}
\bibinfo{author}{Sun, C.} \emph{et~al.}
\newblock \bibinfo{title}{{Single-chip microprocessor that communicates
  directly using light}}.
\newblock \emph{\bibinfo{journal}{Nature}} \textbf{\bibinfo{volume}{528}},
  \bibinfo{pages}{534--538} (\bibinfo{year}{2015}).

\bibitem{Shen2017}
\bibinfo{author}{Shen, Y.} \emph{et~al.}
\newblock \bibinfo{title}{{Deep learning with coherent nanophotonic circuits}}.
\newblock \emph{\bibinfo{journal}{Nature Photonics}}
  \textbf{\bibinfo{volume}{11}}, \bibinfo{pages}{441--446}
  (\bibinfo{year}{2017}).

\bibitem{Bissinger2019}
\bibinfo{author}{Bissinger, J.}, \bibinfo{author}{Ruhstorfer, D.},
  \bibinfo{author}{Stettner, T.}, \bibinfo{author}{Koblm{\"{u}}ller, G.} \&
  \bibinfo{author}{Finley, J.~J.}
\newblock \bibinfo{title}{{Optimized waveguide coupling of an integrated III-V
  nanowire laser on silicon}}.
\newblock \emph{\bibinfo{journal}{Journal of Applied Physics}}
  \textbf{\bibinfo{volume}{125}}, \bibinfo{pages}{243102}
  (\bibinfo{year}{2019}).

\bibitem{Schuster2017}
\bibinfo{author}{Schuster, F.}, \bibinfo{author}{Kapraun, J.},
  \bibinfo{author}{Malheiros-Silveira, G.~N.}, \bibinfo{author}{Deshpande, S.}
  \& \bibinfo{author}{Chang-Hasnain, C.~J.}
\newblock \bibinfo{title}{{Site-Controlled Growth of Monolithic InGaAs/InP
  Quantum Well Nanopillar Lasers on Silicon}}.
\newblock \emph{\bibinfo{journal}{Nano Letters}} \textbf{\bibinfo{volume}{17}},
  \bibinfo{pages}{2697--2702} (\bibinfo{year}{2017}).

\bibitem{Stettner2017}
\bibinfo{author}{Stettner, T.} \emph{et~al.}
\newblock \bibinfo{title}{{Direct Coupling of Coherent Emission from
  Site-Selectively Grown III-V Nanowire Lasers into Proximal Silicon
  Waveguides}}.
\newblock \emph{\bibinfo{journal}{ACS Photonics}} \textbf{\bibinfo{volume}{4}},
  \bibinfo{pages}{2537--2543} (\bibinfo{year}{2017}).

\bibitem{Kanungo2013}
\bibinfo{author}{Kanungo, P.~D.} \emph{et~al.}
\newblock \bibinfo{title}{{Selective area growth of III-V nanowires and their
  heterostructures on silicon in a nanotube template: Towards monolithic
  integration of nano-devices}}.
\newblock \emph{\bibinfo{journal}{Nanotechnology}}
  \textbf{\bibinfo{volume}{24}}, \bibinfo{pages}{225304}
  (\bibinfo{year}{2013}).

\bibitem{Parkinson_phys_apps_NWs}
\bibinfo{author}{Parkinson, P.}
\newblock \bibinfo{title}{{Physics and applications of semiconductor nanowire
  lasers}}.
\newblock In \emph{\bibinfo{booktitle}{Frontiers of Nanoscience}},
  vol.~\bibinfo{volume}{20}, \bibinfo{pages}{389--438}
  (\bibinfo{publisher}{Elsevier}, \bibinfo{year}{2021}).

\bibitem{Shang2022ElectricallyWafers}
\bibinfo{author}{Shang, C.} \emph{et~al.}
\newblock \bibinfo{title}{{Electrically pumped quantum-dot lasers grown on 300
  mm patterned Si photonic wafers}}.
\newblock \emph{\bibinfo{journal}{Light: Science {\&} Applications}}
  \textbf{\bibinfo{volume}{11}}, \bibinfo{pages}{1--8} (\bibinfo{year}{2022}).

\bibitem{Zhu2019ElectricallySi}
\bibinfo{author}{Zhu, S.}, \bibinfo{author}{Shi, B.} \& \bibinfo{author}{Lau,
  K.~M.}
\newblock \bibinfo{title}{{Electrically pumped 15um InP-based quantum dot
  microring lasers directly grown on (001) Si}}.
\newblock \emph{\bibinfo{journal}{Optics Letters}}
  \textbf{\bibinfo{volume}{44}}, \bibinfo{pages}{4566} (\bibinfo{year}{2019}).

\bibitem{Norman2018Perspective:Circuits}
\bibinfo{author}{Norman, J.~C.}, \bibinfo{author}{Jung, D.},
  \bibinfo{author}{Wan, Y.} \& \bibinfo{author}{Bowers, J.~E.}
\newblock \bibinfo{title}{{Perspective: The future of quantum dot photonic
  integrated circuits}}.
\newblock \emph{\bibinfo{journal}{APL Photonics}} \textbf{\bibinfo{volume}{3}},
  \bibinfo{pages}{030901} (\bibinfo{year}{2018}).

\bibitem{DeKoninck2023GaAsLine}
\bibinfo{author}{De~Koninck, Y.} \emph{et~al.}
\newblock \bibinfo{title}{{GaAs nano-ridge laser diodes fully fabricated in a
  300 mm CMOS pilot line}} (\bibinfo{year}{2023}).
\newblock \urlprefix\url{https://doi.org/10.48550/arXiv.2309.04473}.

\bibitem{Wong2021EpitaxiallyLasers}
\bibinfo{author}{Wong, W.~W.}, \bibinfo{author}{Su, Z.}, \bibinfo{author}{Wang,
  N.}, \bibinfo{author}{Jagadish, C.} \& \bibinfo{author}{Tan, H.~H.}
\newblock \bibinfo{title}{{Epitaxially Grown InP Micro-Ring Lasers}}.
\newblock \emph{\bibinfo{journal}{Nano Letters}} \textbf{\bibinfo{volume}{21}},
  \bibinfo{pages}{5681--5688} (\bibinfo{year}{2021}).

\bibitem{Wong_bottom_up_MQW_MRs}
\bibinfo{author}{Wong, W.~W.} \emph{et~al.}
\newblock \bibinfo{title}{{Bottom-up, Chip-Scale Engineering of Low Threshold,
  Multi-Quantum-Well Microring Lasers}}.
\newblock \emph{\bibinfo{journal}{ACS Nano}} \textbf{\bibinfo{volume}{17}},
  \bibinfo{pages}{15065--15076} (\bibinfo{year}{2023}).

\bibitem{Al-Abri2021_ACS}
\bibinfo{author}{Al-Abri, R.}, \bibinfo{author}{Choi, H.} \&
  \bibinfo{author}{Parkinson, P.}
\newblock \bibinfo{title}{{Measuring, Controlling and Exploiting Heterogeneity
  in Optoelectronic Nanowires}}.
\newblock \emph{\bibinfo{journal}{Journal of Physics: Photonics}}
  \textbf{\bibinfo{volume}{3}}, \bibinfo{pages}{022004} (\bibinfo{year}{2021}).

\bibitem{Church2022HolisticNanowires}
\bibinfo{author}{Church, S.~A.} \emph{et~al.}
\newblock \bibinfo{title}{{Holistic Determination of Optoelectronic Properties
  using High-Throughput Spectroscopy of Surface-Guided CsPbBr3 Nanowires}}.
\newblock \emph{\bibinfo{journal}{ACS Nano}} \textbf{\bibinfo{volume}{16}},
  \bibinfo{pages}{9086--9094} (\bibinfo{year}{2022}).

\bibitem{Church2023HolisticDesign}
\bibinfo{author}{Church, S.~A.} \emph{et~al.}
\newblock \bibinfo{title}{{Holistic Nanowire Laser Characterization as a Route
  to Optimal Design}}.
\newblock \emph{\bibinfo{journal}{Advanced Optical Materials}}
  \textbf{\bibinfo{volume}{11}}, \bibinfo{pages}{2202476}
  (\bibinfo{year}{2023}).

\bibitem{Church_data_driven_discovery_2024}
\bibinfo{author}{Church, S.~A.} \emph{et~al.}
\newblock \bibinfo{title}{{Data-Driven Discovery for Robust Optimization of
  Semiconductor Nanowire Lasers}}.
\newblock \emph{\bibinfo{journal}{Laser {\&} Photonics Reviews}}
  \bibinfo{pages}{2401194} (\bibinfo{year}{2024}).

\bibitem{Skalsky2020_ACS}
\bibinfo{author}{Skalsky, S.} \emph{et~al.}
\newblock \bibinfo{title}{{Heterostructure and Q-factor Engineering for
  Low-threshold and Persistent Nanowire Lasing}}.
\newblock \emph{\bibinfo{journal}{Light: Science and Applications}}
  \textbf{\bibinfo{volume}{9}}, \bibinfo{pages}{43} (\bibinfo{year}{2020}).

\bibitem{Yuan2021_Selective}
\bibinfo{author}{Yuan, X.} \emph{et~al.}
\newblock \bibinfo{title}{{Selective area epitaxy of III–V nanostructure
  arrays and networks: Growth, applications, and future directions}}.
\newblock \emph{\bibinfo{journal}{Applied Physics Reviews}}
  \textbf{\bibinfo{volume}{8}}, \bibinfo{pages}{021302} (\bibinfo{year}{2021}).

\bibitem{Parkinson2020}
\bibinfo{author}{Parkinson, P.} \emph{et~al.}
\newblock \bibinfo{title}{{A needle in a needlestack: exploiting functional
  inhomogeneity for optimized nanowire lasing}}.
\newblock In \emph{\bibinfo{booktitle}{Proc. SPIE 11291, Quantum Dots,
  Nanostructures, and Quantum Materials: Growth, Characterization, and Modeling
  XVII}}, \bibinfo{pages}{112910K} (\bibinfo{year}{2020}).
\newblock
  \urlprefix\url{https://www.spiedigitallibrary.org/conference-proceedings-of-spie/11291/2558405/A-needle-in-a-needlestack--exploiting-functional-inhomogeneity-for/10.1117/12.2558405.short}.

\bibitem{Alanis2017}
\bibinfo{author}{Alanis, J.~A.} \emph{et~al.}
\newblock \bibinfo{title}{{Large-scale statistics for threshold optimization of
  optically pumped nanowire lasers}}.
\newblock \emph{\bibinfo{journal}{Nano Letters}} \textbf{\bibinfo{volume}{17}},
  \bibinfo{pages}{4860--4865} (\bibinfo{year}{2017}).

\bibitem{Stettner2016}
\bibinfo{author}{Stettner, T.} \emph{et~al.}
\newblock \bibinfo{title}{{Coaxial GaAs-AlGaAs core-multishell nanowire lasers
  with epitaxial gain control}}.
\newblock \emph{\bibinfo{journal}{Applied Physics Letters}}
  \textbf{\bibinfo{volume}{108}}, \bibinfo{pages}{011108}
  (\bibinfo{year}{2016}).

\bibitem{UFS_Abellana_2023}
\bibinfo{author}{Abellana, D. P.~M.} \& \bibinfo{author}{Lao, D.~M.}
\newblock \bibinfo{title}{{A new univariate feature selection algorithm based
  on the best–worst multi-attribute decision-making method}}.
\newblock \emph{\bibinfo{journal}{Decision Analytics Journal}}
  \textbf{\bibinfo{volume}{7}}, \bibinfo{pages}{100240} (\bibinfo{year}{2023}).

\bibitem{Low_mapping_pareto}
\bibinfo{author}{Low, A.~K.}, \bibinfo{author}{Vissol-Gaudin, E.},
  \bibinfo{author}{Lim, Y.-F.} \& \bibinfo{author}{Hippalgaonkar, K.}
\newblock \bibinfo{title}{{Mapping pareto fronts for efficient multi-objective
  materials discovery}}.
\newblock \emph{\bibinfo{journal}{J Mater Inf}} \textbf{\bibinfo{volume}{3}},
  \bibinfo{pages}{11} (\bibinfo{year}{2023}).

\bibitem{Al-Abri2023Sub-PicosecondFramework}
\bibinfo{author}{Al-Abri, R.} \emph{et~al.}
\newblock \bibinfo{title}{{Sub-Picosecond Carrier Dynamics Explored using
  Automated High-Throughput Studies of Doping Inhomogeneity within a Bayesian
  Framework}}.
\newblock \emph{\bibinfo{journal}{Small}} \textbf{\bibinfo{volume}{19}},
  \bibinfo{pages}{2300053} (\bibinfo{year}{2023}).

\bibitem{Zhang2021AApplications}
\bibinfo{author}{Zhang, F.} \emph{et~al.}
\newblock \bibinfo{title}{{A New Strategy for Selective Area Growth of Highly
  Uniform InGaAs/InP Multiple Quantum Well Nanowire Arrays for Optoelectronic
  Device Applications}}.
\newblock \emph{\bibinfo{journal}{Advanced Functional Materials}}
  \bibinfo{pages}{2103057} (\bibinfo{year}{2021}).

\bibitem{Rudolph2014}
\bibinfo{author}{Rudolph, D.} \emph{et~al.}
\newblock \bibinfo{title}{{Effect of interwire separation on growth kinetics
  and properties of site-selective GaAs nanowires}}.
\newblock \emph{\bibinfo{journal}{Applied Physics Letters}}
  \textbf{\bibinfo{volume}{105}}, \bibinfo{pages}{033111}
  (\bibinfo{year}{2014}).

\bibitem{Noborisaka2005}
\bibinfo{author}{Noborisaka, J.}, \bibinfo{author}{Motohisa, J.} \&
  \bibinfo{author}{Fukui, T.}
\newblock \bibinfo{title}{{Catalyst-free growth of GaAs nanowires by
  selective-area metalorganic vapor-phase epitaxy}}.
\newblock \emph{\bibinfo{journal}{Applied Physics Letters}}
  \textbf{\bibinfo{volume}{86}}, \bibinfo{pages}{1--3} (\bibinfo{year}{2005}).

\bibitem{qNEHVI_qNParEGO}
\bibinfo{author}{Daulton, S.}, \bibinfo{author}{Balandat, M.} \&
  \bibinfo{author}{Bakshy, E.}
\newblock \bibinfo{title}{{Parallel Bayesian Optimization of Multiple Noisy
  Objectives with Expected Hypervolume Improvement}}.
\newblock \emph{\bibinfo{journal}{Advances in Neural Information Processing
  Systems}} \textbf{\bibinfo{volume}{3}}, \bibinfo{pages}{2187--2200}
  (\bibinfo{year}{2021}).

\bibitem{Shahriari2016}
\bibinfo{author}{Shahriari, B.}, \bibinfo{author}{Swersky, K.},
  \bibinfo{author}{Wang, Z.}, \bibinfo{author}{Adams, R.~P.} \&
  \bibinfo{author}{de~Freitas, N.}
\newblock \bibinfo{title}{{Taking the Human Out of the Loop: A Review of
  Bayesian Optimization}}.
\newblock \emph{\bibinfo{journal}{Proceedings of the IEEE}}
  \textbf{\bibinfo{volume}{104}}, \bibinfo{pages}{148--175}
  (\bibinfo{year}{2016}).

\bibitem{GP_Ramussen}
\bibinfo{author}{Rasmussen, C.~E.}
\newblock \bibinfo{title}{{Gaussian Processes in Machine Learning}}.
\newblock In \bibinfo{editor}{Bousquet~O., R.~G., von Luxburg~U.} (ed.)
  \emph{\bibinfo{booktitle}{Advanced Lectures on Machine Learning, ML 2003,
  Lecture Notes in Computer Science}}, vol. \bibinfo{volume}{3176},
  \bibinfo{pages}{63--71} (\bibinfo{publisher}{Springer, Berlin, Heidelberg},
  \bibinfo{year}{2004}).
\newblock
  \urlprefix\url{https://link.springer.com/chapter/10.1007/978-3-540-28650-9_4}.

\bibitem{BoTorch}
\bibinfo{author}{Balandat, M.} \emph{et~al.}
\newblock \bibinfo{title}{{BoTorch: A Framework for Efficient Monte-Carlo
  Bayesian Optimization}}.
\newblock \emph{\bibinfo{journal}{Advances in Neural Information Processing
  Systems}} \textbf{\bibinfo{volume}{33}}, \bibinfo{pages}{21524--21538}
  (\bibinfo{year}{2019}).

\bibitem{EGBO}
\bibinfo{author}{Low, A.~K.} \emph{et~al.}
\newblock \bibinfo{title}{{Evolution-guided Bayesian optimization for
  constrained multi-objective optimization in self-driving labs}}.
\newblock \emph{\bibinfo{journal}{npj Computational Materials}}
  \textbf{\bibinfo{volume}{10}}, \bibinfo{pages}{1--11} (\bibinfo{year}{2024}).

\bibitem{ParEGO}
\bibinfo{author}{Knowles, J.}
\newblock \bibinfo{title}{{ParEGO: A hybrid algorithm with on-line landscape
  approximation for expensive multiobjective optimization problems}}.
\newblock \emph{\bibinfo{journal}{IEEE Transactions on Evolutionary
  Computation}} \textbf{\bibinfo{volume}{10}}, \bibinfo{pages}{50--66}
  (\bibinfo{year}{2006}).

\bibitem{Lopez_Evolutionary_IGD_EMOA}
\bibinfo{author}{Lopez, E.~M.} \& \bibinfo{author}{Coello, C.~A.}
\newblock \bibinfo{title}{{IGD+-EMOA: A multi-objective evolutionary algorithm
  based on IGD+}}.
\newblock In \emph{\bibinfo{booktitle}{2016 IEEE Congress on Evolutionary
  Computation, CEC 2016}}, \bibinfo{pages}{999--1006}
  (\bibinfo{publisher}{Institute of Electrical and Electronics Engineers Inc.},
  \bibinfo{year}{2016}).

\bibitem{Daulton_qEHVI}
\bibinfo{author}{Daulton, S.}, \bibinfo{author}{Balandat, M.} \&
  \bibinfo{author}{Bakshy, E.}
\newblock \bibinfo{title}{{Differentiable Expected Hypervolume Improvement for
  Parallel Multi-Objective Bayesian Optimization}}.
\newblock \emph{\bibinfo{journal}{Advances in Neural Information Processing
  Systems}} \textbf{\bibinfo{volume}{33}}, \bibinfo{pages}{9851--9864}
  (\bibinfo{year}{2020}).

\bibitem{McKay_LHS}
\bibinfo{author}{McKay, M.~D.}, \bibinfo{author}{Beckman, R.~J.} \&
  \bibinfo{author}{Conover, W.~J.}
\newblock \bibinfo{title}{{A Comparison of Three Methods for Selecting Values
  of Input Variables in the Analysis of Output from a Computer Code}}.
\newblock \emph{\bibinfo{journal}{Technometrics}}
  \textbf{\bibinfo{volume}{21}}, \bibinfo{pages}{239} (\bibinfo{year}{1979}).

\bibitem{PCA_dim_red}
\bibinfo{author}{Velliangiri, S.}, \bibinfo{author}{Alagumuthukrishnan, S.} \&
  \bibinfo{author}{Thankumar~Joseph, S.~I.}
\newblock \bibinfo{title}{{A Review of Dimensionality Reduction Techniques for
  Efficient Computation}}.
\newblock \emph{\bibinfo{journal}{Procedia Computer Science}}
  \textbf{\bibinfo{volume}{165}}, \bibinfo{pages}{104--111}
  (\bibinfo{year}{2019}).

\end{thebibliography}

\end{document}